\documentclass[prl,showpacs,preprintnumbers,amsmath,amssymb]{revtex4}

 \def\be{\begin{equation}}
 \def\ee{\end{equation}}
 \def\bes{\begin{eqnarray}}
 \def\ees{\end{eqnarray}}

 \def\2{\frac{1}{2}}
 \def\4{\frac{1}{4}}

\catcode`\@=11
\def\@citex[#1]#2{%
\if@filesw \immediate \write \@auxout {\string \citation {#2}}\fi
\@tempcntb\m@ne \let\@h@ld\relax \def\@citea{}%
\@cite{%
  \@for \@citeb:=#2\do {%
    \@ifundefined {b@\@citeb}%
      {\@h@ld\@citea\@tempcntb\m@ne{\bf ?}%
      \@warning {Citation `\@citeb ' on page \thepage \space
undefined}}%
      {\@tempcnta\@tempcntb \advance\@tempcnta\@ne%
      \@tempcntb\number\csname b@\@citeb \endcsname \relax%
      \ifnum\@tempcnta=\@tempcntb 
it
        \ifx\@h@ld\relax%
          \edef \@h@ld{\@citea\csname b@\@citeb\endcsname}%
        \else%
          \edef\@h@ld{\ifmmode{-}\else--\fi\csname
b@\@citeb\endcsname}%
        \fi%
      \else
        \@h@ld\@citea\csname b@\@citeb \endcsname%
        \let\@h@ld\relax%
      \fi}%
    \def\@citea{,\penalty\@highpenalty\,}%
  }\@h@ld
}{#1}}

\def\@citeb#1#2{{[#1]\if@tempswa , #2\fi}}
\def\@citeu#1#2{{$^{#1}$\if@tempswa , #2\fi }}
\def\@citep#1#2{{#1\if@tempswa , #2\fi}}

\begin{document}
\preprint{UTHET-08-0901}

\title{Cosmology from an AdS Schwarzschild black hole via holography}

\author{Pantelis S.~Apostolopoulos$^1$}
 \email{pantelis.apost@uib.es}
\author{George Siopsis$^2$}
 \email{siopsis@tennessee.edu}
\author{Nikolaos Tetradis$^3$}
\email{ntetrad@phys.uoa.gr}
\affiliation{%
$^1$Departament de F\'isica, Universitat de les Illes Balears, 
Carretera Valldemossa Km 7.5, E-07122 Palma de Mallorca, Spain \\\\\\
$^2$Department of Physics and Astronomy,
The University of Tennessee,
Knoxville, TN 37996 - 1200, USA \\\\\\
$^3$Department of Physics,
University of Athens,
University Campus,
Zographou, 157 84, Greece}
\date{\today}%

\begin{abstract}

We derive the equations of cosmological evolution from an AdS Schwarzschild black hole 
via holographic renormalization with appropriate boundary conditions.

 \end{abstract}

\maketitle


The effect of extra dimensions demanded by string theory on cosmology has been 
extensively investigated. In a popular scenario \cite{rs}, one considers a three-brane 
within a higher-dimensional bulk. The cosmological evolution of the brane  
\cite{binetruy} is equivalent to its
motion within the bulk space \cite{kraus}. 
The bulk may be occupied by a black hole, or a more complicated (or unkown) 
solution to the Einstein field equations \cite{apostol}. The second possibility 
seems more appropriate 
for a cosmological setup and allows for energy exchange between the brane and
the bulk \cite{exchange}. 

Without the brane-world assumption, the connection 
of the AdS/CFT correspondence to cosmology is elusive.
Our aim is to understand how cosmological evolution emerges in the context of 
AdS/CFT \cite{bib1}.
We shall show that the equations of cosmological evolution emerge via holographic 
renormalization \cite{Skenderis}, even when one starts
from a static AdS Schwarzschild black hole, provided the boundary conditions are chosen 
appropriately.

In general, one starts with a solution to the bulk Einstein equations with a 
negative cosmological constant and proceeds to compute the properties of the 
dual gauge theory at strong coupling on the AdS boundary.
One usually fixes the geometry of the boundary by adopting Dirichlet boundary conditions. 
However, in order to obtain cosmological evolution 
the boundary geometry must remain dynamical.
Changing the boundary conditions in order to accommodate a dynamical boundary metric 
may lead to fluctuations of the bulk metric which are not normalizable \cite{bib2}. 
It was recently shown that such fears are unfounded, and the boundary geometry can 
be dynamical if one correctly introduces boundary counterterms needed in order to 
cancel infinities \cite{bib3}.

We shall concentrate on an AdS Schwarzschild black hole in five dimensions, 
which is a solution to the Einstein field equations with a 
negative cosmological constant 
($\Lambda_5 = -6l^2$),
\be\label{eqein} R_{AB} = \frac{2}{l^2} g_{AB} \ , \ee
where $A,B = 0,1,2,3,4$.
The metric can be written in static coordinates as
\be\label{eqmetric} ds^2 = -f(r) dt^2 + \frac{dr^2}{f(r)} 
+ r^2 d\Omega_k^2 \  , \ \ \ \ f(r) = r^2+k - \frac{\mu}{r^2} \ , \ee
where $k=+1,0,1$
for spherical, flat and hyperbolic horizons, respectively. We set $l =1$ for simplicity.

The Hawking temperature and mass of the hole are, respectively,
\be\label{eqTM} T_H = \frac{2r_+^2 + k}{2\pi r_+} \ , \ \ \ \ 
M = \frac{3V_k}{16\pi G_5} r_+^2 (r_+^2 + k) \ ,\ee
where $V_k$ is the volume of the three-dimensional space $\Sigma_k$ spanned by 
$\Omega_k$, $r_+$ is the radius of the horizon and $G_5$ 
is Newton's constant in the bulk.

The Einstein equations (\ref{eqein}) are obtained by varying the bulk action 
$I_{\mathcal{M}}$. This is a five-dimensional Einstein-Hilbert action on
$\mathcal{M}$ with a cosmological term. It also includes a Gibbons-Hawking boundary 
term, as well as boundary counterterms needed to render the system finite.
In addition to the black hole solution in the bulk, the variation of the action in 
general yields a boundary term
\be\label{eqvar} \delta I_{\mathcal{M}} = \frac{1}{2} \int_{\partial \mathcal{M}} 
d^4 x \sqrt{-\det g^{(0)}} T_{\mu\nu}^{(CFT)} \delta g^{(0)\mu\nu} \ , \ee
where $g_{\mu\nu}^{(0)}$ ($\mu,\nu = 0,1,2,3$) is the boundary metric and 
$T_{\mu\nu}^{(CFT)}$ the stress-energy tensor of the dual conformal field 
theory (CFT) on the AdS boundary $\partial\mathcal{M}$.

If one adopts Dirichlet boundary conditions that fix $g_{\mu\nu}^{(0)}$, 
the additional term (\ref{eqvar}) vanishes. As we are interested in 
keeping $g_{\mu\nu}^{(0)}$ dynamical, following \cite{bib3}, 
we shall adopt {\em mixed} boundary conditions instead. 
To define them, we shall introduce
a boundary action consisting of two terms,
\be\label{eq5} I_{\partial\mathcal{M}} = 
I_{\partial\mathcal{M}}^{(EH)} + I_{\partial\mathcal{M}}^{(matter)} \ . \ee
The first term is the Einstein-Hilbert action in four dimensions
\be I_{\partial\mathcal{M}}^{( EH)} = - \frac{1}{16\pi G_4} 
\int_{\partial\mathcal{M}} d^4 x \sqrt{-\det g^{(0)}} (\mathcal{R} - 2\Lambda_4) \ , 
\ee
where $G_4$ ($\Lambda_4$) is Newton's constant in 
the four-dimensional boundary and $\mathcal{R}$ the four-dimensional 
Ricci scalar (constructed from $g_{\mu\nu}^{(0)}$).
The second term is an unspecified action for matter fields,
\be\label{eqmat4} I_{\partial\mathcal{M}}^{( matter)} = 
\int_{\partial \mathcal{M}} d^4 x \sqrt{-\det g^{(0)}} \mathcal{L}^{(matter)} \ . \ee
The matter fields reside on the boundary and have no bulk duals.

To the variation of the bulk action (\ref{eqvar})
we must now add the variation of the new boundary action 
$\delta I_{\partial\mathcal{M}}$, given by (\ref{eq5}), 
under a change in the boundary metric.
Demanding that the sum vanish
\be \delta I_{\mathcal{M}} +\delta I_{\partial\mathcal{M}} =0 \ee
leads to two possibilities: 
{\em (a)} Dirichlet boundary conditions, i.e. fixed $g_{\mu\nu}^{(0)}$, 
or {\em (b)} {\em mixed} boundary conditions:
\be\label{eqcosmo} \mathcal{R}_{\mu\nu} - \frac{1}{2} g_{\mu\nu}^{(0)} \mathcal{R} 
- \Lambda_4 g_{\mu\nu}^{(0)} = 8\pi G_4 \left( T_{\mu\nu}^{(CFT)} 
+ T_{\mu\nu}^{(matter)} \right) \ .\ee
The latter, which we shall adopt, are nothing but the 
Einstein field equations in four dimensions.
Moreover, the variation of the boundary matter action (\ref{eqmat4}) under a
change in the matter fields leads to 
the standard four-dimensional matter field equations.

In \cite{bib3} it was shown that a general form of 
$I_{\partial\mathcal{M}} [ g^{(0)}]$ leads to a sensible theory with normalizable 
metric fluctuations. Matter fields were not considered. However, the framework 
may be extended to include boundary matter fields. If one 
integrates over them in the path integral, an effective action is obtained
as a functional of the boundary metric $g^{(0)}$. For this 
effective action, the discussion in \cite{bib3} is applicable.

In order to understand the AdS/CFT correspondence, it is useful to write the metric in 
terms of Fefferman-Graham coordinates \cite{fg}. Define $z$ through
\be \frac{dz}{z} = -\frac{dr}{\sqrt{f(r)}} \ee
which gives (with an appropriate integration constant)
\be z^4 = \frac{16}{k^2+4\mu} \ \frac{r^2 + \frac{k}{2} - r\sqrt{f(r)}}{r^2 
+ \frac{k}{2}+ r\sqrt{f(r)}} \ee
This equation may be inverted to give
\be\label{eq4a} r^2 = \frac{\alpha + \beta z^2 + \gamma z^4}{z^2} \ee
where
\be\label{eq4x}
\alpha = 1 \ , \ \
\beta = - \frac{k}{2} \ , \ \ \gamma = \frac{k^2+4\mu}{16}\ . \ee
The metric (\ref{eqmetric}) reads
\be\label{eqmetric1} ds^2 = \frac{1}{z^2} 
\left[ dz^2 - \frac{\left( 1- \gamma z^4 \right)^2}{1+\beta z^2 + \gamma z^4} dt^2 
+ \left( 1+ \beta z^2 +\gamma z^4 \right) d\Omega_k^2 \right] \ . \ee

The energy of the dual conformal field theory on the AdS boundary is found 
through holographic renormalization. For a metric in the form
\be\label{eq2} ds^2 = \frac{1}{z^2} \left[ dz^2 + g_{\mu\nu} dx^\mu dx^\nu \right] \ee
where
\be g_{\mu\nu} = g_{\mu\nu}^{(0)} + z^2 g_{\mu\nu}^{(2)} 
+ z^4 g_{\mu\nu}^{(4)} + \dots \ee
the stress-energy tensor of the CFT is \cite{Skenderis}
\be\label{eq3a} \langle T_{\mu\nu}^{(CFT)} \rangle =
\frac{1}{4\pi G_5} 
\left\{ g^{(4)} - \frac{1}{4} g^{(2)} g^{(0)} g^{(2)}
+ \frac{1}{4} \mathrm{tr} (g^{(2)}(g^{(0)})^{-1})\, g^{(2)}
- \frac{1}{8} \left[ \left( \mathrm{tr} (g^{(2)}(g^{(0)})^{-1}) \right)^2 
- \mathrm{tr} (g^{(2)} (g^{(0)})^{-1})^2  \right] g^{(0)} 
\right\}_{\mu\nu} \ . \ee
Applying this general expression to our metric (\ref{eqmetric1}), 
we obtain the energy density and pressure, respectively,
\be\label{eq3} \langle T_{tt}^{(CFT)} \rangle = 3\langle T_{ii}^{(CFT)} \rangle = 
\frac{3\gamma}{4\pi G_5}  \ee
on the static Einstein Universe $\mathbb{R}\times \Sigma_k$ with metric
\be\label{eqmetric0} ds_0^2 = g_{\mu\nu}^{(0)} dx^\mu dx^\nu = -dt^2 + d\Omega_k^2 \ .\ee
Notice that the total energy $E=\langle T_{tt}^{(CFT)} \rangle V_k$ is larger 
than the mass of the black hole (eq.~(\ref{eqTM})) by a constant (Casimir energy) in the 
case of a curved horizon ($k\ne 0$). The two quantities agree for flat horizons ($k=0$). 
We shall show that the additional piece can be understood by a change of the vacuum 
state from the Minkowski to the conformal vacuum. (Note that the curved metrics 
(\ref{eqmetric0}) can be conformally mapped on the Minkowski space.)


For general cosmological applications, instead of the static boundary considered 
above (with metric (\ref{eqmetric0})), we need a boundary with the 
form of a Robertson-Walker spacetime
\be\label{eqmetricb} ds_0^2 = g_{\mu\nu}^{(0)} dx^\mu dx^\nu = 
-d\tau^2 + a^2(\tau) d\Omega_k^2 \ ,\ee
on which to apply holographic renormalization. 
To this end, we need to choose a different foliation away from the black hole, 
consisting of hypersurfaces whose metric is asymptotically of the form (\ref{eqmetricb}).
Therefore, we need to make a change of coordinates 
$(t,r) \to (\tau, z)$ and bring the black-hole metric (\ref{eqmetric}) 
in the form
%
\be\label{eqmetric3} ds^2 = \frac{1}{z^2} \left[ dz^2 
- \mathcal{N}^2(\tau,z) d\tau^2 + \mathcal{A}^2 (\tau,z) d\Omega_k^2 \right]\ , \ee
where $\mathcal{N}(\tau,z)\to 1$ and $\mathcal{A}(\tau,z)\to a(\tau)$ 
as we approach the boundary $z=0$.
Comparison with the static case (eq.~(\ref{eq4a})) suggests the {\em ansatz}
\be \mathcal{A}^2 = \alpha(\tau) + \beta(\tau) z^2 + \gamma (\tau) z^4 \ , \ee
where $\alpha(\tau), \beta(\tau), \gamma(\tau)$ are functions to be determined.

The function $\mathcal{N}$ is constrained by the $\tau z$ component of the 
Einstein equations (\ref{eqein}) to be of the form
\be \mathcal{N} = \frac{\dot{\mathcal{A}}}{\delta (\tau)} \ . \ee
Agreement with the boundary metric (\ref{eqmetricb}) then fixes
\be \alpha (\tau) = a^2 (\tau) \ \ , \ \ \ \ \delta (\tau) = \dot a (\tau) \ . \ee
The diagonal components of the Einstein equations collectively yield
\be \beta = - \frac{\dot a^2 + k}{2} \ .\ee
The rest of the Einstein equations are satisfied provided
\be \beta \dot\beta = 2(\dot\alpha\gamma + \alpha\dot\gamma)\ , \ee
which may be integrated to give
\be\label{eqgam} \gamma = \frac{(\dot a^2 +k)^2 +4\mu}{16a^2} \ , \ee
where we fixed the integration constant by comparing with the 
static case (eq.~(\ref{eq4x})).

Thus, the metric (\ref{eqmetric3}) is uniquely specified. 
It agrees with the black hole metric (\ref{eqmetric}) provided
\bes\label{eqsys1} \frac{(r')^2}{f(r)} - f(r) (t')^2 &=& z^{-2} \nonumber\\
\frac{r' \dot r}{f(r)} - f(r) t'\dot t &=& 0 \nonumber\\
\frac{\dot r^2}{f(r)} - f(r) \dot t^2 &=& - \mathcal{N}^2 z^{-2} \nonumber\\ r 
&=& \mathcal{A}z^{-1} \ . \ees
The last equation fixes $r(\tau,z)$. Two of the other three equations may 
then be used to determine the derivatives $\dot t$ and $t'$. We obtain
\be\label{eqsys2} \dot t = -\frac{\dot A r'}{f \dot a} \ \ , 
\ \ \ \ t' = -\frac{\dot a}{zf} \ .\ee
In fact, these expressions satisfy all three equations.
One can then verify the consistency of the system (\ref{eqsys1}) by calculating the 
mixed derivative $\dot t'$ using each of the two equations (\ref{eqsys2}), 
and showing that the two expressions match. 
Upon integration we obtain a unique function $t(\tau,r)$, up to an irrelevant constant.
General explicit expressions are unwieldy and will not be reported here. For
example, for pure AdS space in Poincar\'e coordinates ($\mu = 0$, $k=0$), we obtain
\be t(\tau ,z) = -\frac{2\dot a z^2}{4a^2 - {\dot a^2} z^2} 
+ \int^\tau \frac{d\tau'}{a(\tau')} \ , \ee
so that at the boundary ($z=0$) the coordinate $t$ reduces to conformal 
time $\int^\tau {d\tau'}/{a(\tau')}$, while it 
receives corrections as we move into the bulk.
This is generally the case.
However, general explicit expressions are not needed in order to extract 
physical results because we already know the explicit form of the metric 
in the new coordinates (eqs.~(\ref{eqmetric3}) - (\ref{eqgam})).

The stress-energy tensor of the dual CFT on the cosmological boundary (\ref{eqmetricb}) 
is determined via holographic renormalization (eq.~(\ref{eq3a})). 
We obtain the energy density and pressure, respectively,
\bes\label{eq4} \langle ( T^{(CFT)} )_{\tau \tau} \rangle 
&=& \frac{3}{64\pi G_5} \ \frac{(\dot a^2+k)^2 + 4\mu}{a^4} 
\nonumber\\ 
\langle ( T^{(CFT)} )_{i}^i \rangle &=&  \frac{(\dot a^2 + k)^2+4\mu 
-4a\ddot a (\dot a^2 + k)}{64\pi G_5a^4} \ ,\ees
where no summation over $i$ is implied.
($i$ can be chosen in any 
spatial direction due to isotropy.)
We deduce the conformal anomaly
\be g^{(0)\mu\nu} \langle T_{\mu\nu}^{(CFT)} \rangle = 
- \frac{3\ddot a (\dot a^2 + k)}{16\pi G_5a^3} \ .\ee

The above results can also be derived directly from the standard expressions 
for the energy and pressure of a gauge-theory plasma in Minkowski space
through an entirely four-dimensional calculation.
To this end, we observe that the Robertson-Walker metric (\ref{eqmetricb}) 
is conformally equivalent to the flat Minkowski metric.
The vacuum expectation value (\ref{eq4}) is calculated in the {\em conformal vacuum}. 
We may instead calculate the VEV in the {\em Minkowski vacuum}, 
in which case we obtain the static plasma result.
The two VEVs are related through \cite{bib4}
\be\label{eq33} \langle T_{\mu\nu}^{(CFT)}\rangle \Big|_{RW} = 
\frac{1}{a^4} \langle T_{\mu\nu}^{(CFT)} \rangle \Big|_{Minkowski}
 + \mathbf{a} H_{\mu\nu}^{(1)} +\mathbf{b} H_{\mu\nu}^{(3)} \ ,\ee
with
\bes H_{\mu\nu}^{(1)} &=& 2(\nabla_\nu\nabla_\mu - g_{\mu\nu}^{(0)} \nabla^2 )\mathcal{R} 
-\frac{1}{2} g_{\mu\nu}^{(0)} \mathcal{R}^2 + 2\mathcal{R}\mathcal{R}_{\mu\nu} 
\nonumber\\
H_{\mu\nu}^{(3)} &=& \frac{1}{12} \mathcal{R}^2 \delta_\mu^\nu - \mathcal{R}^{\rho\sigma} 
\mathcal{R}_{\rho\mu\sigma}^{\ \ \ \ \nu} \ .\ees
The curvature tensor $\mathcal{R}_{\ \ \nu\rho\sigma}^\mu$ is
derived from the Robertson-Walker metric $g_{\mu\nu}^{(0)}$ of eq. (\ref{eqmetricb}).
The coefficients $\mathbf{a}$ and $\mathbf{b}$ are related to the conformal anomaly 
and depend on the field content of the CFT. 
For a theory with $n_B$ spin-0 bosons, $n_F$ fermions and $n_V$ vector fields, 
we have
\be \mathbf{a} = - \frac{n_B + 3n_F - 18n_V}{1080(4\pi)^2} \ \ , \ \ \ \ 
\mathbf{b} = \frac{n_B+\frac{11}{2} n_F + 62 n_V}{180 (4\pi)^2} \ .\ee
For the $\mathcal{N} = 4$ $U(N)$ super-Yang-Mills theory, 
we have $n_B = 6N^2$, $n_F =4N^2$ and $n_V=N^2$.
Therefore
\be \mathbf{a} = 0 \ \ , \ \ \ \ \mathbf{b} = \frac{N^2}{32\pi^2}\ . \ee
After some algebra, eq.~(\ref{eq33}) is seen to agree with (\ref{eq4}), with
$G_5 \sim N^{-2}$.

The temperature on the boundary may also be understood by comparing with the case of 
a static plasma. For a plasma in the static Einstein Universe (\ref{eqmetric0}), 
the temperature coincides with the Hawking temperature $T_H$ of the 
black hole (\ref{eqTM}). Since the RW metric (\ref{eqmetricb}) is conformally 
equivalent to (\ref{eqmetric0}), the conformal factor being $a^2$, the Euclidean 
proper time period of thermal Green functions in the RW metric scales as $a$. 
As a result, the temperature $T$ of the Universe (inversely proportional to the period) 
scales as $a^{-1}$. It coincides with $T_H$ when $a=1$.

Finally, the boundary conditions (\ref{eqcosmo}) yield the equation of cosmological 
evolution
\be H^2 + \frac{k}{a^2} -\frac{\Lambda_4}{3}
= \frac{1}{16\pi G_5} \left[ \left(  H^2 + \frac{k}{a^2} 
\right)^2 + \frac{4\mu}{a^4} \right] + \frac{8\pi G_4}{3} \rho \ ,\ee
where $\rho = T_{00}^{(matter)}$ is the energy density of four-dimensional 
``ordinary'' matter (without a bulk dual) and we have introduced the Hubble parameter
$H = {\dot a}/{a}$. This equation has the expected form, reflecting the
conformal anomaly and the presence of a radiative energy component whose energy
density scales $\sim a^{-4}$.
 
We have discussed how the equations of cosmological
evolution can be obtained in the context of the AdS/CFT correspondence. Two 
essential elements are necessary: a) the choice of appropriate boundary conditions, so
that the boundary metric becomes dynamical, and b) the derivation of the gravity 
solution in terms of coordinates such that the boundary metric has the
Robertson-Walker form. We have demonstrated that the procedure correctly
reproduces the expected four-dimensional cosmological behavior, starting from 
a static AdS Schwarzschild five-dimensional solution. The challenge for the future
is to repeat the procedure for gravity duals of more realistic theories. This 
requires the deviation from conformal invariance and the presence of additional
fields. This procedure may lead to the understanding of the non-perturbative
aspects of cosmological phase transitions. For example, the deconfinement 
phase transition could be discussed starting from a holographic QCD model
\cite{gursoy}.

\section*{Acknowledgments}
We wish to thank E.~Kiritsis, G.~Comp\`ere and D.~Marolf for useful discussions. 
P.S.A. was supported by the Spanish Ministerio 
de Educaci\'{o}n {y} Ciencia through the Juan de la Cierva program and also 
through the research grants FPA-2007-60220, PCTIB-2005-GC2-06 
(Ministerio de Educaci\'{o}n {y} Ciencia) and PROGECIB-2A 
(Conselleria Economia, Hisenda i Innovaci\'{o} del Govern Illes Balears).
G.~S. was 
supported in part by the Department of Energy under grant DE-FG05-91ER40627.
N.~T. was supported in part by the EU Marie Curie Network ``UniverseNet'' 
(HPRN--CT--2006--035863).

\end{document}